\documentclass[twocolumn,
reprint,
bibnotes,
amsmath,amssymb,
aps,
prl,
]{revtex4-1}

\usepackage{graphicx}
\usepackage{dcolumn}
\usepackage{bm}
\usepackage{relsize}

\begin{document}
	
	
	\title{Effective Bounds on Network-Size for Anti-phase Synchronization }
	
	\author{George Vathakkattil Joseph}
	\email{george.vathakkattiljoseph@ucdconnect.ie}
	\author{Vikram Pakrashi}%
	
	\affiliation{%
		Dynamical Systems and Risk Lab, University College Dublin, Ireland
	}%

	\date{\today}
	
	\begin{abstract}
	We consider anti-phase synchronization of coupled oscillators using the Stuart-Landau model and explore its relative infrequency in occurrence compared to in-phase synchronization. We report effective limits in number of oscillators which can anti-phase synchronize for general configurations of real-world networks. We link anti-phase synchronization to the Ising model and consequently to combinatorial optimization problems, thereby explaining experimentally observed limits in self-organization of natural systems. We illustrate this using the Steiner-tree problem. 
	\end{abstract}
	
	\pacs{Valid PACS appear here}
	\maketitle

	
	Coupled oscillators and phenomena associated with them have attracted considerable attention recently not only due to the behavioural richness they exhibit but also due to their generality in capturing the essential dynamics of multiple real-world systems (eg. lasers\cite{laser1,laser2}, Josephson junctions\cite{josephson}, neural networks\cite{neuralnet}, ecological systems\cite{ecology}, etc. \cite{boccaletti2018synchronization}). Synchronization of phase oscillators has probably been the most studied phenomenon in this burgeoning field, spanning from the classical insight from Huygen's pendulum clocks and the simplicity of Kuramoto's model 3 centuries later\cite{strogatz2000kuramoto}. Kuramoto's phase oscillator model is a versatile approximation and often serves as a popular model for studying synchronization of such oscillators. However, many natural systems exhibit self-stabilizing amplitude dynamics as well, which motivated the study of the Stuart-Landau (SL) model. The SL oscillator is the simplest nonlinear extension of the harmonic oscillator and can also be viewed as the discrete form of the complex Ginzburg Landau equation \cite{ginzburglandau}. This makes the SL model arguably the most widely applicable model to study coupled oscillator dynamics\cite{rohm2018bistability}. 
	
	One of the simplest forms of synchronization is \textit{phase locking} where the phase difference between oscillators becomes and remains a constant. It is common to refer to a stronger notion of synchronization where the phases of the oscillators become and remain identical regardless of their initial conditions. This is usually termed as in-phase synchronization and we use the same definition. We also note that some form of attractive coupling is essential for the oscillators to 'pull' others in the network to a common phase.
	
	Another form of synchronization, in-fact the first form as observed by Huygens,\cite{boccaletti2018synchronization} is anti-phase synchronization, where the phases of oscillators are separated by a phase difference of $\pi$. From the simple case of two oscillators, it is easy to deduce that some form of repulsive coupling would be required in order to achieve anti-phase synchronization \cite{kimpattern}. 
	
	Although the two forms of synchronization may appear to be symmetric with change in coupling direction, this is not the case except for the simple 2-oscillator example. The anti-phase synchronized state becomes unstable quickly as the number of oscillators is increased. Tsimring et al.\cite{tsimring2005repulsive} considered the case of identical phase oscillators, with a homogeneous global coupling and showed that for purely repulsive coupling, anti-phase synchronization or rather synchronization of any kind is not possible. For non-identical oscillators, an empirical upper bound of 3 was found for the number of phase oscillators that can synchronize. Therein, the phase-locking definition of synchronization was used, however, we use the definition that phases should be \textit{equal} for in-phase (IP) synchronization and separated by 180 degrees for antiphase (AP) synchronization (Fig.1). 
	
		\begin{figure}[b]
		\resizebox{\columnwidth}{!}{%
			\includegraphics{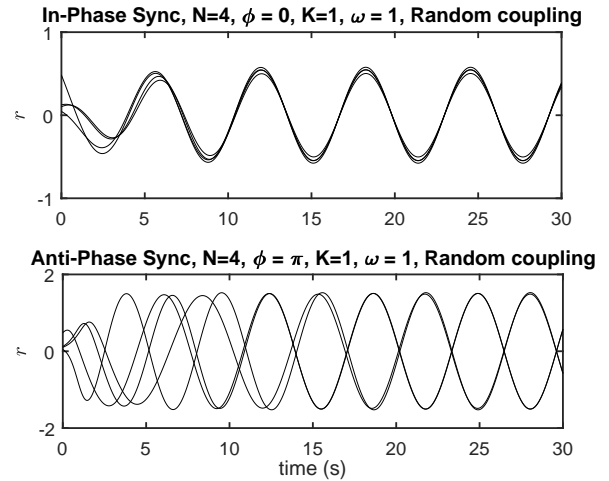} }
		\caption{4 identical oscillators, in-phase and antiphase synchronizing for random coupling}
		\label{fig:1}       
	\end{figure}

	Allowing for amplitude dynamics using the SL equation relaxes the observed cap on oscillator count slightly as we show in this Letter (N=4 case in Fig.1), but nevertheless, the probability of achieving AP synchronization for arbitrary networks quickly reduces with network size. The limitation on the number of oscillators that can AP synchronize is a fundamental result that can explain the expected size of anti-phase patterns in nature. In biological networks where repulsive coupling is common \cite{repulsive_bio_1,repulsive_bio_2,repulsive_bio_3}, this result can place an upper bound on the expected size of a cluster in which an anti-phase pattern may occur. The generality of the SL model implies that numerous equivalent systems are constrained to the same general limit. 
	
	Consider the Ising model, we show that the $\pm1$ binary spins at equilibrium are equivalent to the $0$ and $\pi$ phases of an AP synchronized oscillator network. For an arbitrary Ising Hamiltonian, the above result would limit the number of globally coupled spins that can self-organize (to the optimal ground-state) to a finite number. Rephrasing the Ising model as an optimization problem, if the number of entities in a network is greater than the limit $n$, for arbitrary coupling, the chances of the system finding the global minimum by self-organizing is negligible. Many such optimization problems of practical importance can be expressed as Ising spin-glasses, ranging from protein folding\cite{protein1,protein2} and memory to collective decision making in economics\cite{economic} and social sciences\cite{social}. A list of such NP-Complete and NP-Hard problems mapped to the Ising model is available\cite{lucas2014ising}, one of which is the Steiner tree problem \cite{steiner}: For N points on a plane, the goal of the problem is to connect them by lines of minimum total length forming a tree where every point is connected. 
	
	Aaronson and others popularized an elegant soap bubble experiment\cite{aaronson2005guest} to showcase the formation of a Steiner tree when two glass plates with N pegs connecting the two are dipped and raised from soapy water. The result was that with 3 or 4 pegs, the optimal solution is usually found, but as the peg count increases, the solution gets stuck in a sub-optimal configuration. The result of this Letter provides an explanation to the observed limit. 
				
	We begin by analyzing a general network of $N$ SL oscillators given by the equation:
	\begin{equation}
	\dot{z_i}= (\lambda + i\omega - \gamma|z_i|^2)z_i + \frac{K}{N}e^{i\phi} \sum\limits_{j=1}^{N}A_{ij}[z_j-z_i]
	\end{equation}
	where $\lambda$ determines the Hopf bifurcation (at $\lambda=0$), $\omega$ is the natural frequency, $Re(\gamma)$ determines if the bifurcation is super- or sub-critical, $Im(\gamma)$ is similar to the hardness of the spring and causes an amplitude-phase coupling, $K$  is the coupling constant, $N$ is the number of oscillators and $\phi$ is the coupling phase. The coupling between oscillators is given by the coupling (adjacency) matrix $A$ where each element $A_{ij} \in (0 \ 1], A_{ii}=0$. If $A_{ij}=1$ $ \forall i\neq j$, then we have homogeneous coupling, else heterogeneous. 

	Our approach is to attempt at AP synchronizing as many oscillators as possible using idealized cases to obtain global limits on $N$ if any, and then relax the restrictions to arrive at expected limits for common configurations in real-world applications. For each case, we use random numerical trials to estimate probability of the outcome. We use $(\lambda, K, \gamma) = (1,3,1)$ unless specified otherwise.

	To achieve AP synchronization, from a qualitative perspective, the network should be able to self-organize into two groups as widely separated as possible from the other group, but with oscillators within the group as close as possible to each other. This makes repulsive coupling a prerequisite and we set the coupling phase $\phi = \pi$. This also brings in a distinction between even and odd $N$-networks. Homogeneous even networks will be easier to synchronize as 2 groups can easily form. For homogeneous odd networks, any permutation into two groups would never be stable. However, for heterogeneous odd networks, the possibility of having two clusters exists. In real-world networks, where perfect homogeneity is improbable, both odd and even networks may form AP synchronized patterns. Fig2a) shows how for identical oscillators, homogeneous coupling prevents AP sync almost always beyond 2 oscillators, whereas random coupling, where elements of the coupling matrix are drawn from a uniform distribution in (0 1), allows more oscillators to AP synchronize. 
	
	Although the general classical network models would adopt random graphs, it has been shown that symmetry is ubiquitous in real world networks \cite{macarthur2008symmetry}. Homogeneous coupling is obviously symmetric to any degree, but networks with heterogeneous coupling can have degrees of symmetry as well. The simplest case is when the oscillator coupling strengths are mirrored about an axis, forming two symmetric groups. The coupling matrix $A_{ij}$ is bisymmetric in this case. Allowing mirror symmetry enhances AP synchronization probability, P(AP-Sync) slightly for higher N as shown in Fig2a. Regardless, P(AP-Sync) falls below 0.5 by N=6.  
	
	Creating an axis of separation in the graph by setting the cross-diagonal elements of $A_{ij}$ close to $0$, has a strong impact on increasing AP synchronization. When coupled with symmetry (or without), this enhances the ability to AP synchronize making $P(AP-Sync)\approx 0.50$ at $N=8$. 
	
	Our last parameter of interest that can significantly affect P(AP-Sync) is Amplitude-Phase coupling, given by $Im(\gamma)$. Amplitude-phase coupling (APC), also known as anisochronicity, has been shown to be of importance in various applications, such as shear in fluid dynamics, susceptibility variation of gain medium in laser cavities, etc \cite{bohm2015amplitude}. If each oscillator has strong APC ($|Im(\gamma)|=50$), then the network can AP synchronize for any even N (Fig2c) in networks with homogeneous coupling. For random cases, the outcome is similar to the cases without APC. (Note that when APC is added to the oscillators, the coupling phase $\phi$ for IP and AP synchronization is shifted from 0 and $\pi$ by $\pi/2$ to $\pi/2$ and $3\pi/2$ as explained in \cite{rohm2018bistability}.)
	
	Finally, considering the improbability of exactly identical oscillators in real-world networks, we allow a normal spread in natural frequency about 1 with a standard deviation of 0.05 (Fig2b). This takes P(AP-Sync) for N$>$6 below 0.1 even for symmetric or cross-diagonal weakened cases.

		\begin{figure*}
		\resizebox{2\columnwidth}{!}{%
			\includegraphics{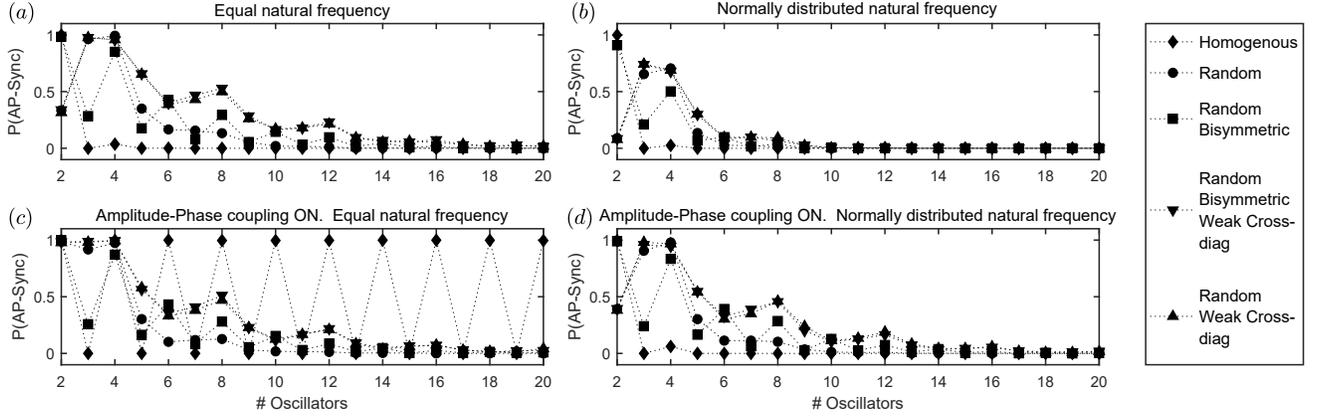} }
		\caption{Probability of anti-phase synchronization for (a) identical natural frequencies $\omega =1$ (b) normally distributed natural frequencies with standard deviation = 0.05 and mean = 1(c) Amplitude-phase coupling $Im(\gamma) =50$, identical natural frequencies $\omega =1$ (d)Amplitude-phase coupling $Im(\gamma) =50$, normally distributed natural frequencies with standard deviation = 0.05 and mean = 1}
		\label{fig:2}       
	\end{figure*}


	Recently, research into a novel computational method called the coherent Ising machine has shown that the global minimum of the Ising Hamiltonian can be reached by placing an analog version of the Hamiltonian in a double-well potential that imposes binary constraints to the analog spins\cite{leleu2017combinatorial}. For the simplified Ising Hamiltonian
	\begin{equation}
	H(x) = \sum\limits_{j=1}^{N}J_{ij}x_jx_i
	\end{equation}
	combined with the bistable potential $V_b(x_i) = -(1/2)\alpha x_i^2 + (1/4)x_i^4$ as 
	\begin{equation}
	V = \sum_{i} [V_b(x_i)] + \epsilon H(x)
	\end{equation}
	the analog spin dynamics is given by
	\begin{equation}
	\dot{x_i}= \alpha x_i -x_i^3 + \epsilon \sum_{i\neq j}J_{ij}x_j
	\label{eqn_ising}
	\end{equation}
	At steady state, assuming homogeneous amplitudes $\dot{x}=0$ and $\sigma_i = x_i/|x_i|$ giving the sign of the spin,
	\begin{equation}
	\alpha - x^2 + \epsilon \sum_{i\neq j}J_{ij}\sigma_i \sigma_j =0
	\label{steady}
	\end{equation}
	\begin{equation}
	\Rightarrow x^2 =\alpha - \frac{2\epsilon}{N}H
	\end{equation}	
	For $\alpha$ above a threshold, the first non-zero steady state gives the minimal value of $H$. (A detailed explanation is available in \cite{leleu2017combinatorial})
	
	If Eqn.\ref{eqn_ising} is generalized to the complex plane and assuming a natural frequency $\omega$ for the spins, the equation becomes
	\begin{equation}
	\dot{z_i}= (\alpha +i\omega -|z_i|^2)z_i + \epsilon \sum_{i\neq j}J_{ij}z_j
	\end{equation}
	The terms before the sum represent SL oscillators with $\alpha = \lambda$. At steady state, assuming uniform amplitude, 
	\begin{equation}
	\alpha +i\omega - |z|^2 + \epsilon \sum_{i\neq j}J_{ij}\hat{z_i} \hat{z_j} =0
	\end{equation}
	where $\hat{z_i}$ represent the unit vectors in the complex plane. This is equivalent to Eqn.\ref{steady} if $\hat{z}\equiv\sigma$, which is only satisfied if $\hat{z}\in\{-1,1 \}$. Now,
	\begin{equation}
		\sum_{i\neq j}J_{ij}\hat{z_i} \hat{z_j} = \sum_{i\neq j}J_{ij} \cos(\theta_j - \theta_i)
		\label{energy}
	\end{equation}
	which gives $\theta_j - \theta_i \in \{0,\pi\}$  ie. the anti-phase synchronized case. The coupling matrix $A_{ij}$ encodes interaction terms $J_{ij}$ of the Hamiltonian. (The equivalence in Eqn.\ref{energy} may be alternatively considered as the equivalence of the energy functions of the Ising model and the XY model.)  
	
	\begin{figure}[h]
		\resizebox{\columnwidth}{!}{%
			\includegraphics{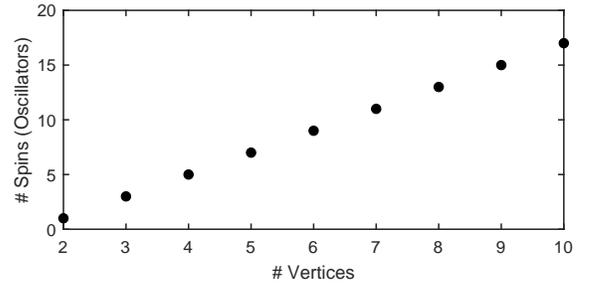} }
		\caption{Number of spins (y) required to simulate a Steiner tree of (x) vertices}
		\label{fig:3}       
	\end{figure}
	
	Consider the Steiner tree problem as an example of a combinatorial optimization problem which can be formulated as an Ising problem: a Steiner tree problem with $V$ vertices and $E$ possible edges can be expressed as an Ising Hamiltonian where each binary variable decides the existence of a possible edge in the tree.  The maximum number of Steiner points in a Steiner tree is $V - 2$ and maximum number of edges in a graph is $V - 1$ giving the number of spins, $s(V)=2V-3$ for the maximal case. The spins required for various sizes of graphs are shown in Fig.3. The number of spins required exceeds $6$ as $V>4$, implying that the probability of a system finding an optimal Steiner tree by self organizing for $V>4$ is small whereas for trees with 3 or 4 vertices, it usually finds the optimal solution, agreeing with the experiments \cite{aaronson2005guest}.
	
	The Steiner tree problem serves as a lucid example due to the availability of the simple yet insightful experiment with soap bubbles. Other important combinatorial optimization problems such as partitioning, packing, and colouring problems are bound by the same limits which we did not explore here. The result presented in this Letter may hence give a physical explanation for limits found elsewhere in nature. 

	\bibliography{main}
	
\end{document}